\documentclass[journal,10pt,twocolumn,draftclsnofoot]{IEEEtran}
\usepackage{setspace}
\usepackage{cite}
\usepackage{amsmath,amssymb,amsfonts}
\usepackage{algorithmic}
\usepackage{graphicx}
\usepackage{graphics}
\usepackage{textcomp}

\usepackage{subcaption}
\usepackage{multirow}
\usepackage[inline]{enumitem}
\usepackage[table,x11names]{xcolor}
\definecolor{yellow}{rgb}{1,1,0}
\definecolor{darkblue}{rgb}{0.4,0.7,1}
\definecolor{lightblue}{rgb}{0.7,0.9,1}

\usepackage{fancyhdr}
\begin{document}

\title{Towards a Design Framework for TNN-Based Neuromorphic Sensory Processing Units
 %
}

\author{\IEEEauthorblockN{Prabhu Vellaisamy\textsuperscript{1} and John Paul Shen\textsuperscript{2}}\\
\IEEEauthorblockA{\textit{Electrical and Computer Engineering} \\
\textit{Neuromorphic Computer Architecture Lab}\\
\textit{Carnegie Mellon University}\\
pvellais@andrew.cmu.edu\textsuperscript{1} jpshen@cmu.edu\textsuperscript{2}}
}


\maketitle
\thispagestyle{fancy}

\begin{abstract}
Temporal Neural Networks (TNNs) are spiking neural networks that exhibit brain-like sensory processing with high energy efficiency. This work presents the ongoing research towards developing a custom design framework for designing efficient application-specific TNN-based \textit{Neuromorphic Sensory Processing Units} (NSPUs). This paper examines previous works on NSPU designs for UCR time-series clustering and MNIST image classification applications. Current ideas for a custom design framework and tools that enable efficient software-to-hardware design flow for rapid design space exploration of application-specific NSPUs while leveraging EDA tools to obtain post-layout netlist and power-performance-area (PPA) metrics are described. Future research directions are also outlined.
\end{abstract}


\section{Introduction}
Deep Neural Networks (DNNs) have achieved impressive performance on various sensory applications such as computer vision, speech, and object recognition \cite{lecun2015deep}. However, their computing demands have increased exponentially \cite{openai} and are unsustainable given their computational and economic costs \cite{thompson2020computational}. Temporal Neural Networks (TNNs) \cite{smith2018space, smith2017space, smith2020temporal} are a special class of Spiking Neural Networks (SNNs) that use precise spike timings to represent and process information, mimicking the brain's \textit{temporal} computational paradigm. 

TNNs employ simple feed-forward networks operating on spike-timing relationships, unlike DNNs that use high-dimensional tensor processing. 
One distinguishing feature of TNNs is temporal encoding, in which information is represented by the relative timing of the spikes, requiring just a single spike to encode one value. 
Furthermore, TNNs use biologically plausible Spike Timing Dependent Plasticity (STDP) local learning rules, enabling them to perform online continuous learning, while DNNs use compute-intensive backpropagation, which separates training and testing phases. These two distinctive attributes of TNNs make them the most ``neuromorphic" or brain-like.

Chaudhary et al.\cite{chaudhary2021unsupervised} demonstrate the efficacy of using TNNs for unsupervised time-series clustering, which can be employed for edge-native applications such as anomaly detection and healthcare monitoring. A microarchitectural framework for \textit{direct} CMOS implementation of TNNs has been proposed in \cite{nair2021online} and successfully used to implement key TNN building blocks: the neurons, columns, synapses, and the STDP learning rules.

This paper presents the current research in developing an automated software-to-hardware design framework for designing compelling sensory processing units based on TNNs, namely \textit{Neuromorphic Sensory Processing Units} (NSPUs). 
Targeted applications for NSPUs include diverse sensory signal processing for the edge, mobile, and IoT devices. Due to the potential of extreme energy efficiency, NSPUs can enable always-on, on-device, autonomous, and highly distributed processing.

We envision a design framework for NSPUs that incorporates: (1) a scalable microarchitecture model for implementing application-specific NSPUs, and
(2) a suite of supporting tools, including a PyTorch-based software simulator to perform rapid design space explorations and automated tools for RTL-to-GDSII flow to produce post-layout power-performance-area (PPA) metrics for NSPUs.

\section{Background \& Initial Results}
Due to their neuromorphic attributes, NSPUs exhibit extreme energy efficiency and are ideal for always-on edge-native processing devices. Initial results in \cite{nair2021online} report post-synthesis evaluation of the TNN column implementations in 45 nm and demonstrate the efficacy of using TNNs for MNIST digit classification using supervised learning. A \textit{large} column configuration of 1024 synapses and 16 output neurons consists of 1.7M gates, with the area and power footprint of 1.65 mm\textsuperscript{2} and 7.96 mW, which is less than 1\% of the area and power budget of mobile SoCs \cite{nair2021online}.

\cite{chaudhary2021unsupervised} illustrates the efficacy of using TNNs for unsupervised time-series clustering of 36 UCR benchmark datasets and demonstrates that TNN designs either outperform or are competitive to the state-of-the-art algorithms. Furthermore, hardware complexity analysis is performed by doing standard technology scaling to achieve predictive 7nm results from 45 nm post-synthesis results for the designs employed for the 36 datasets. Ongoing work on developing and incorporating custom TNN column-based macro cells, named \textit{TNN7} \cite{nair2022tnn7}, shows promising PPA improvements ($\sim$17\% less power, $\sim$16\% faster, and $\sim$27\% less area) over standard-cell based designs - the largest TNN column used in \cite{chaudhary2021unsupervised} with 6,750 synapses consumes just 0.054 mm\textsuperscript{2} area, 39 $\mu$W power, and 28.14 ns computation time. These results establish that developing NSPU designs capable of online unsupervised and supervised learning has enormous potential in edge-native sensory processing applications.
\section{NSPU Microarchitecture Framework}

\begin{figure}[t]
\centering
\includegraphics[width=1\columnwidth, height = 6cm]{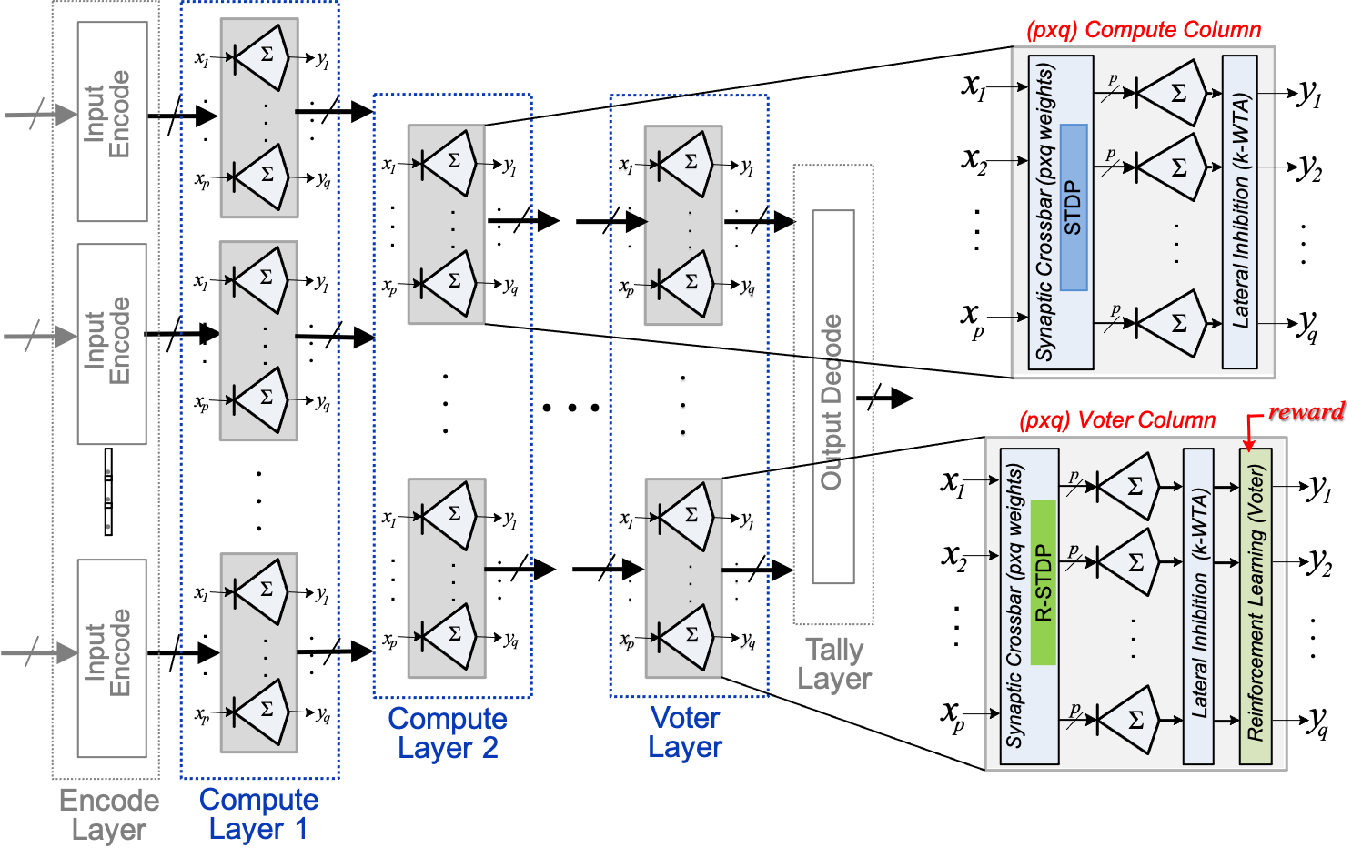} 
\caption{Hierarchical TNN Organization}
\label{fig:neuron_uarch}
\end{figure}

Fig. \ref{fig:neuron_uarch} depicts a high-level hierarchical architecture of TNNs, where
each neuron has \textit{p} synaptic inputs and one output, and each synapse carries a local synaptic weight that is updated based on the relative timing of the incoming spike to that synapse and the outgoing spike from the neuron body. The STDP learning rule determines the synaptic weight update, and through it, a neuron \textit{learns} an input feature by adapting the weight to match the corresponding input pattern. A detailed description of TNN microarchitecture is provided in \cite{nair2021online}.

The fundamental building block for NSPUs is a \textit{column}; we denote a column configuration of \textit{p}x\textit{q} as containing \textit{q} excitatory neurons and a synaptic crossbar connecting \textit{p} inputs to the \textit{q} neurons via \textit{p}x\textit{q} synapses. Larger TNN designs can be constructed by stacking multiple columns to form a multi-column layer, and also by cascading multiple layers into a large multi-layer TNN. Multi-layer TNNs have been utilized for MNIST digit recognition. \cite{smith2020temporal} has shown a 4-layer TNN with 3.096M synapses can achieve 1\% error rate on MNIST.

\section{Automated Toolsuite for NSPUs}

The ongoing research aims to develop a custom design framework for efficient and automated mapping of NSPU designs from software to hardware, facilitating rapid design space exploration and yielding post-layout PPA values from place-\&-route. The entire framework will consist of the following - (i) \textit{TNNSim}, a functional TNN simulator implemented in PyTorch \cite{paszke2019pytorch}, (ii) \textit{TNNGen} - a script suite for the Cadence toolchain for performing automated RTL-to-GDSII flow to generate predictive 7 nm PPA values, (iii) the ASAP7 \cite{clark2016asap7} standard cell library, (iv) a custom macro library containing TNN functional units developed as an extension to the ASAP7 library that leads to optimized TNN column design, and (v) a Python library function, such as PyVerilog \cite{takamaeda2015pyverilog} or MyHDL \cite{decaluwe2004myhdl}, to convert Python codes to RTL. 

\begin{figure}[t]
\centering
\includegraphics[width=1\columnwidth, height = 5.2 cm]{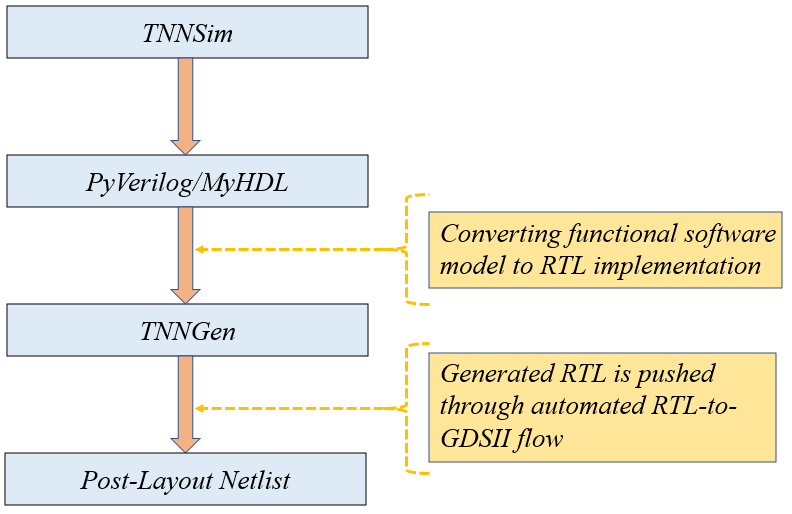} 
\caption{Custom NSPU Design Framework Flow}
\label{fig:framework}
\end{figure}

The design framework, depicted in Fig. \ref{fig:framework}, will allow high-level users to experiment with constructing application-specific NSPU designs while getting accurate hardware complexity analysis results in predictive 7 nm. A typical user will implement PyTorch NSPU models for specific sensory processing applications by invoking TNNSim's built-in library of TNN functional modules, finetune the architectures to achieve the best performing models and then choose to initiate the hardware flow. Once initiated, PyVerilog or MyHDL is leveraged to map the NSPU functional models to RTL. This flow is more optimal than high-level synthesis (HLS) flows as it allows direct Verilog or VHDL scripting and outputs desired optimal RTL implementations. Consequently, TNNGen is invoked, and automated RTL-to-GDSII flow occurs, involving the Cadence tools like Genus for synthesis, Innovus for place-\&-route, and Voltus for accurate signoff power numbers.  

\section{Future Research Directions}
A complete NSPU design framework, incorporating TNNSim and TNNGen, is being developed. Researchers and developers can use this framework and its suite of tools to explore the design and implementation of a wide range of application-specific NSPUs, targeting IoT, edge, and mobile devices. This framework can also facilitate the design of specialized NSPUs for edge-native acceleration of AI/ML workloads that consume less than 10 mW of power. 
Furthermore, this framework can serve as a foundation for implementing arbitrary space-time functional units \cite{smith2018space} in the near future.

\bibliographystyle{IEEEtranS}
\bibliography{refs}

\end{document}